\documentclass[conference]{IEEEtran}
\IEEEoverridecommandlockouts
\usepackage{cite}
\usepackage{amsmath,amssymb,amsfonts}
\usepackage{algorithmic}
\usepackage{graphicx}
\usepackage{textcomp}
\usepackage{xcolor}
\def\BibTeX{{\rm B\kern-.05em{\sc i\kern-.025em b}\kern-.08em
    T\kern-.1667em\lower.7ex\hbox{E}\kern-.125emX}}

\DeclareMathOperator*{\argmin}{arg\,min}
\usepackage{mathtools}

\usepackage{steinmetz}

\usepackage[ruled,norelsize]{algorithm2e}
\makeatletter
\newcommand{\removelatexerror}{\let\@latex@error\@gobble}
\makeatother
\usepackage{caption}
\usepackage{subcaption}

\begin{document}

\title{Rate-Splitting Multiple Access for Joint Radar-Communications with Low-Resolution DACs}


\author{\IEEEauthorblockN{Onur Dizdar$^1$, Aryan Kaushik$^2$, Bruno Clerckx$^1$, Christos Masouros$^2$}
\IEEEauthorblockA{$^1$Department of Electrical and Electronic Engineering, Imperial College London, United Kingdom.\\
$^2$Department of Electronic and Electrical Engineering, University College London (UCL), United Kingdom. \\
Emails: \{o.dizdar, b.clerckx\}@imperial.ac.uk, \{a.kaushik, c.masouros\}@ucl.ac.uk}}

\maketitle

\begin{abstract}
In this paper, we introduce the design of a multi-antenna Joint Radar-Communication (JRC) system with Rate Splitting Multiple Access (RSMA) and low resolution Digital-to-Analog Converter (DAC) units. Using RSMA, the communication messages are split into private and common parts, then precoded and quantized before transmission. We use a problem formulation to design the JRC system with RSMA and low resolution DACs by maximizing communication sum-rate and the proximity of the resulting JRC waveform to an optimal radar beampattern under an average transmit power constraint. We solve the joint sum-rate maximization and beampattern error minimization problem using Alternating Direction Method of Multipliers (ADMM) method. The numerical results show that RSMA achieves a significantly higher sum-rate compared to Space Division Multiple Access (SDMA) while providing the same Normalized Mean Square Error (NMSE) for the designed radar beampattern.
\end{abstract}


\IEEEpeerreviewmaketitle

\section{Introduction}
Joint Radar-Communication (JRC) systems share the common spectral and hardware resources to efficiently use the limited radio frequency spectrum \cite{blissACCESS2017, fanTCOM2020, hassanienSP2019}. Some existing JRC systems focus on the single-antenna radar systems \cite{tangTAES2019, zhengTSP2018}, which can be upgraded to multi-antenna systems to obtain improved radar performance \cite{stoica2007}. In this direction, the use of multiple antennas has been discussed in the existing literature to exploit more degrees of freedom in joint sensing and communication systems \cite{fanWCL2017, cuiSPAWC2018, fanTSP2018}. For the multi-antenna JRC systems, managing the interference among the communication users and radar targets efficiently has significant importance. 

Rate-Splitting Multiple Access (RSMA) is a multiple access technique for multi-antenna systems that relies on linearly precoded RS at the transmitter and successive interference cancellation (SIC) at the receivers \cite{clerckxCM2016, clerckx2018}. 
RSMA manages interference in a flexible and robust manner by partially decoding interference and treating the remaining interference as noise. The RSMA technique outperforms existing multiple access schemes such as Space Division Multiple Access (SDMA) and Non-Orthogonal Multiple Access (NOMA) \cite{clerckx2018,clerckxTC2016, clerckxWCL2020}. Therefore, RSMA is a natural candidate to be employed in JRC system to manage the interference efficiently.

In addition to the interference management, reducing the hardware complexity and designing a power efficient system is also of high importance in JRC systems. In \cite{aryanTGCN2019, aryanTGCN2020, vlachosRS2020}, the authors provide energy efficient frameworks using low-resolution Digital-to-Analog Converters (DACs) and Analog-to-Digital Converters (ADCs) for communication-only Multiple-Input Multiple-Output (MIMO) systems.
The authors provide radio-frequency chain optimization for MIMO based JRC systems in \cite{aryanICC2021}. 
Low resolution sampling can be implemented to further save power consumption, as discussed in \cite{aryanGcom2019} for a communication-only system. References \cite{kumariSSP2018, kumariIcassp2020} suggest the use of low resolution ADCs for the JRC systems. However, the use of low resolution DAC sampling has not been widely exploited for multi-antenna JRC systems. Similarly, the impact of low resolution DACs on RSMA has not been investigated.

In this paper, we investigate RSMA for JRC systems with low resolution DACs under a total transmit power budget. RSMA was first studied for JRC in \cite{xuICCW2020} without considering low resolution DACs.
We formulate an optimization problem for the communication rate and the proximity of the JRC waveform to a designed radar beampattern by involving the impact of low resolution DAC distortion and a total transmit power budget.   
The resulting non-convex problem is solved using Alternating Direction Method of Multipliers (ADMM) method. 
We analyze the sum-rate and Normalized Mean Square Error (NMSE) performance, all with respect to the variation in the number of DAC quantization bits. We consider a total transmit power budget to be shared among the transmit precoders and the DACs. Simulation results demonstrate that the maximum sum-rate is achieved by a number of quantization bits which is smaller than the maximum number allowed under the total transmit power budget. Furthermore, we show that RSMA achieves a significantly higher sum-rate than SDMA for all considered numbers of quantization bits.

\textit{Notation:} Vectors and matrices are denoted by bold lowercase letters and bold uppercase letters, respectively; $|.|$ and $||.||_{2}$ are the absolute value of a scalar and $l_{2}$-norm of a vector, respectively. The vector $\mathbf{a}^{H}$ is the Hermitian transpose of a vector $\mathbf{a}$. $\mathcal{CN}(0,\sigma^{2})$ denotes the Circularly Symmetric Complex Gaussian distribution with zero mean and variance $\sigma^{2}$. The matrix $\mathbf{I}_{n}$ denotes the $n$-by-$n$ identity matrix. 
The operator $\mathrm{Diag}(\mathbf{X}_{1}, \ldots, \mathbf{X}_{K})$ builds a matrix $\mathbf{X}$ by placing the matrices $\mathbf{X}_{1}$, $\ldots$, $\mathbf{X}_{K}$ diagonally and setting all other elements to zero. 
The operator $\mathrm{diag}(\mathbf{X})$ builds a vector $\mathbf{x}$ from the diagonal elements of $\mathbf{X}$.

\section{RSMA and Quantization Based System Model}
\subsection{System Model}
\begin{figure}[t]
\centering
    \includegraphics[width=0.5\textwidth, trim=200 120 280 80,clip]{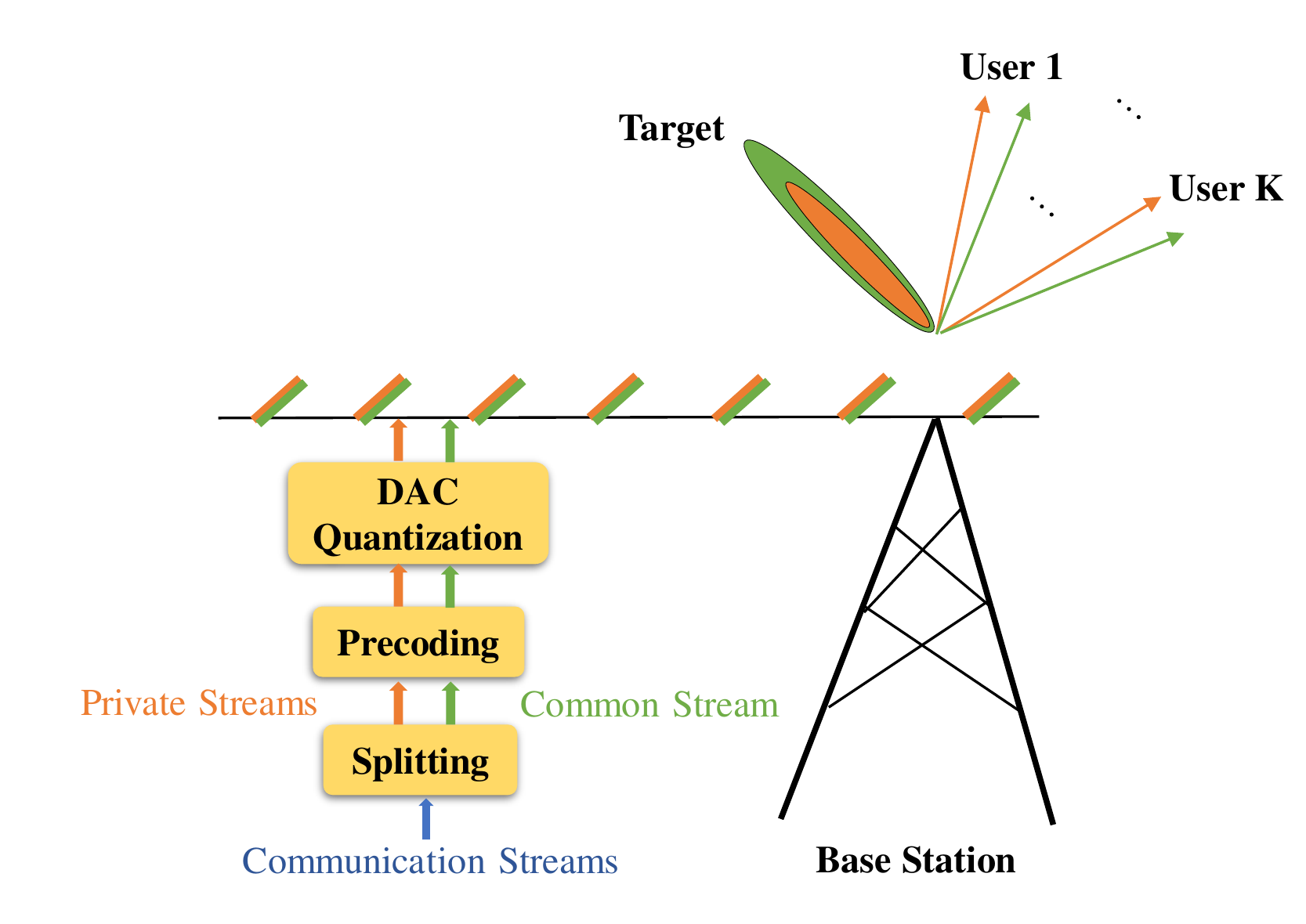}
		\caption{JRC system with RSMA and Low Resolution DACs.}
		\label{fig:system}
		\vspace{-0.5cm}
\end{figure}

We consider a JRC system consisting of one transmitter with $N_{t}$ transmit antennas serving $K$ single-antenna users indexed by $\mathcal{K}=\left\lbrace1,\ldots ,K \right\rbrace $ while detecting a single target. 
We employ RSMA to perform multiple access communications. Fig.~\ref{fig:system} shows the RSMA based JRC system with low resolution DAC quantization.

RSMA splits the user messages into common and private parts, encodes the common parts of the user messages into a common stream to be decoded by all receivers, encodes the private parts of the user messages into private streams and superposes them in a non-orthogonal manner.
We consider that the messages intended for the communication users, $W_{k}$, are split into common and private parts, i.e., $W_{c,k}$ and $W_{p,k}$, $\forall k\in \mathcal{K}$. The common parts of the messages, $W_{c,k}$, are combined into the common message $W_{c}$. The common message $W_{c}$ and the private messages $W_{p,k}$ are independently encoded into streams $s_{c}$ and $s_{k}$, $\forall k\in \mathcal{K}$, respectively. Linear precoding is applied to all streams with $\mathbf{P}=\left[\mathbf{p}_{c},\mathbf{p}_{1},\ldots,\mathbf{p}_{K}\right] $, where $\mathbf{p}_{c}$, $\mathbf{p}_{k} \in\mathbb{C}^{n_t}$ are the precoders for the common stream and the private stream of the user-$k$, respectively. The communication signal at the transmitter is expressed as
\begin{align}
	\mathbf{x}=\mathbf{p}_{c}s_{c}+\sum_{k=1}^{K}\mathbf{p}_{k}s_{k}.
	\label{eqn:transmit_signal}	
\end{align}

We assume that the streams have unit power, so that \mbox{$\mathbb{E}\left\lbrace \mathbf{s}\mathbf{s}^{H}\right\rbrace =\mathbf{I}$}, where \mbox{$\mathbf{s}=[s_{c}, s_{1}, \ldots, s_{K}]$}. We assume that $s_{c}$ and $s_{k}$, $\forall k\in\mathcal{K}$, are chosen independently from a Gaussian alphabet for theoretical analysis. The signal received by user-$k$ user is written as
\begin{align}
 \mathbf{y}_{k}&=\mathbf{h}_{k}^{H}\mathbf{x}+z_{k}, \quad \forall k \in \mathcal{K}, 
 \label{eqn:receive_signal}	
\end{align}
where \mbox{$\mathbf{h}_{k} \in \mathbb{C}^{n_{t}}$} is the channel vector and \mbox{$z_{k} \sim \mathcal{CN}(0,\sigma_{n}^{2})$} is the Additive White Gaussian Noise (AWGN) component for user-$k$. 

The detection of the messages is carried out using the SIC algorithm. The common stream is detected first to obtain the common message estimate $\hat{W}_{c}$ by treating the private streams as noise. The common stream is then reconstructed using $\hat{W}_{c}$ and subtracted from the received signal. The remaining signal is used to detect the private messages $\hat{W}_{p,k}$. Finally, the estimated message for user-$k$, $\hat{W}_{k}$, is obtained by combining $\hat{W}_{c,k}$ and $\hat{W}_{p,k}$. 


We note here that SDMA or conventional multiuser linear precoding is a subscheme of \eqref{eqn:transmit_signal} that is obtained when we do not allocate any power to the common stream $s_{c}$ and encode $W_{k}$ into $s_{k}$. Hence, the sequel also holds for SDMA by simply turning off the common stream \cite{clerckxCM2016,clerckx2018,clerckxTC2016,clerckxWCL2020}.

\subsection{Quantization Model}
We consider a linear model approximation for the quantization noise of the DACs, as also discussed in \cite{kaushik_2019}. We define the uniform scalar quantizer function $Q(x)$ for an RF chain as
\begin{align}
Q(u)\approx \delta u+\epsilon,
\end{align}
where the parameter $\delta$ represents the quantization resolution of $b$ bits and is expressed in terms of $b$ as
\begin{align}
    \delta=\sqrt{1-\frac{\pi\sqrt{3}}{2}2^{-2b}}.
    \label{eqn:delta}
\end{align} 
The quantization noise \mbox{$\epsilon \sim \mathcal{CN}(0,\sigma_{e}^{2})$} and the input signal $u$ are independent. The quantization noise variance is expressed as $\sigma_{e}=\delta^{2}(1-\delta^{2})^{2}$. 

We assume that each RF chain has identical number of quantization bits. Substituting \eqref{eqn:transmit_signal}, the quantized transmitted signal $\mathbf{x}$ is written as
\begin{align}
Q(\mathbf{x})\approx \delta\mathbf{x}+\boldsymbol{\epsilon},
\label{eqn:quantization}
\end{align}
where \mbox{$\boldsymbol{\epsilon} \sim \mathcal{CN}(0,\sigma_{e}^{2}\mathbf{I}_{n_{t}})$} and independent of $\mathbf{x}$. The power consumption of each active DAC is proportional to the number of quantization bits. The power consumption model is expressed as
\begin{align}
P(\delta)=\left[ P_\textrm{DAC}\sqrt{\frac{\pi\sqrt{3}}{2(1-\delta^{2})}} \right], 
\label{eqn:power}
\end{align}
with $P_\textrm{DAC}$ being the power consumption coefficient. In the next section, we discuss the sum-rate maximization and radar beampattern design problem which includes the impact of low resolution DACs, and present the ADMM-based solution.

\section{Sum-Rate Maximization and Transmit Radar Beampattern Design}

\subsection{Problem Formulation}
\label{sec:problem}
We can express the transmitted signal for RSMA under the quantization effects of the DACs from \eqref{eqn:transmit_signal} and \eqref{eqn:quantization} as  
\begin{align}
 \tilde{\mathbf{x}}=Q(\mathbf{x})=\delta(\mathbf{p}_{c}s_{c}+\sum_{k=1}^{K}\mathbf{p}_{k}s_{k})+\boldsymbol{\epsilon}=\delta\mathbf{P}\mathbf{s}+\boldsymbol{\epsilon}.
 \label{eqn:transmit_signal_quant}
\end{align}
Using \eqref{eqn:receive_signal} and \eqref{eqn:transmit_signal_quant}, the received RSMA signal at the $i$-th user under DAC quantization error is written as
\begin{align}
\tilde{\mathbf{y}}_{i}&=\mathbf{h}_{i}^{H}\tilde{\mathbf{x}}\hspace{-0.07cm}+\hspace{-0.07cm}z_{i}= \mathbf{h}_{i}^{H}\delta\mathbf{p}_{c}s_{c}\hspace{-0.07cm}+\hspace{-0.07cm}\sum_{k=1}^{K}\mathbf{h}_{i}^{H}\delta\mathbf{p}_{k}s_{k} \hspace{-0.07cm}+\hspace{-0.07cm}\underbrace{\mathbf{h}_{i}^{H}\boldsymbol{\epsilon}+z_{i}}_{\eta_{i}},
\label{eqn:receive_signal_quant}	
\end{align}
where \mbox{$\eta_{i} \sim \mathcal{CN}(0,\sigma^{2}_{\eta,i})$} and $\sigma^{2}_{\eta,i}=\sigma_{e}^{2}\mathbf{h}_{i}^{H}\mathbf{h}_{i}+\sigma_{n}^{2}$.
As it can be observed from the expression \eqref{eqn:receive_signal_quant}, the effects of DAC quantization is reflected by a multiplicative factor $\delta$ on the precoders and an increased noise variance.

We express the received SINR values for the common and private streams as 
\begin{align}
\gamma_{c,k}(\mathbf{P}, \delta, \sigma_{e})&=\frac{\delta^{2}|\mathbf{h}_{k}^{H}\mathbf{p}_{c}|^{2}}{\sigma^{2}_{\eta,i}+\delta^{2}\sum_{i \in \mathcal{K}}|\mathbf{h}_{k}^{H}\mathbf{p}_{i}|^{2}}, \nonumber \\
&=\frac{|\mathbf{h}_{k}^{H}\mathbf{p}_{c}|^{2}}{\frac{\sigma^{2}_{\eta,i}}{\delta^{2}}+\sum_{i \in \mathcal{K}}|\mathbf{h}_{k}^{H}\mathbf{p}_{i}|^{2}}, \label{eqn:sinr_1}\\
\gamma_{k}(\mathbf{P}, \delta, \sigma_{e})&=\frac{\delta^{2}|\mathbf{h}_{k}^{H}\mathbf{p}_{k}|^{2}}{\sigma^{2}_{\eta,i}+\delta^{2}\sum_{i \in \mathcal{K}, i \neq k}|\mathbf{h}_{k}^{H}\mathbf{p}_{i}|^{2}} \nonumber \\
&=\frac{|\mathbf{h}_{k}^{H}\mathbf{p}_{k}|^{2}}{\frac{\sigma^{2}_{\eta,i}}{\delta^{2}}+\sum_{i \in \mathcal{K}, i \neq k}|\mathbf{h}_{k}^{H}\mathbf{p}_{i}|^{2}}. \label{eqn:sinr_2}
\end{align}


We extend the optimization problem in \cite{xuICCW2020} to obtain the optimal precoders for JRC with DAC quantization.
Our objective function jointly maximizes the sum-rate and minimizes the radar beampattern error \mbox{$\sum_{m=1}^{M}|\alpha P_{d}(\theta_{m})-\mathbf{a}^{H}(\theta_{m}) \mathbf{R}\mathbf{a}(\theta_{m})|^{2}$}, with $\theta_{m}$ representing the $m$-th azimuth angle grid in degrees, for $m=\left\lbrace 1,2, \ldots, M\right\rbrace $, $P_{d}(\theta_{m})$ representing the desired beampattern amplitude at $\theta_{m}$ and $\mathbf{R}=\mathbb{E}\left\lbrace \tilde{\mathbf{x}}\tilde{\mathbf{x}}^{H} \right\rbrace$ is the covariance matrix of the signal at the transmitter \cite{stoica_2007}, calculated as
\begin{align}
    \mathbf{R}=\mathbb{E}\left\lbrace (\delta\mathbf{P}\mathbf{s}+\boldsymbol{\epsilon})(\delta\mathbf{P}\mathbf{s}+\boldsymbol{\epsilon})^{H} \right\rbrace=\delta^{2}\mathbf{P}\mathbf{P}^{H}+\sigma_{e}^{2}\mathbf{I}_{N_{t}}.
    \label{eqn:R}
\end{align}
The cross-correlation terms do not appear in \eqref{eqn:R} due to the random quantization error being zero-mean and independent of the transmit symbols. Substituting \eqref{eqn:R} into the error metric, the radar beampattern error is obtained as \mbox{$\sum_{m=1}^{M}|\alpha P_{d}(\theta_{m})-\delta^{2}\mathbf{a}^{H}(\theta_{m}) \mathbf{P}\mathbf{P}^{H}\mathbf{a}(\theta_{m})-\sigma_{e}^{2}N_{t}|^{2}$}.
Note that the effect of the random quantization noise $\boldsymbol{\epsilon}$ on the radar beampattern error is independent of the precoder matrix. 
The transmit steering vector $\mathbf{a}(\theta_{m})$ is defined as \mbox{$\mathbf{a}(\theta_{m})=[1,e^{j2\pi sin(\theta_{m})d}, \ldots,e^{j2\pi (N_{t}-1)sin(\theta_{m})d} ]$}, where $d$ is the normalized distance between adjacent array elements with respect to wavelength.

We use the SINR expressions \eqref{eqn:sinr_1} and \eqref{eqn:sinr_2} and the radar beampattern error above into the considered formulation to capture the effects of quantization. Furthermore, we include the power consumption of each DAC, $P(\delta)$, in the total transmitter power budget to be shared with the power allocated to the precoders. The resulting problem formulation is given as
 \begin{subequations}
	\begin{alignat}{3}
	\max_{\alpha,\mathbf{c}, \mathbf{P}}&     \quad  \sum_{k\in\mathcal{K}}(C_{k}+R_{k}(\mathbf{P}, \delta, \sigma_{e})) \nonumber \\ &-\lambda\sum_{m=1}^{M}|\alpha P_{d}(\theta_{m})-\delta^{2}\mathbf{a}^{H}(\theta_{m}) \mathbf{P}\mathbf{P}^{H}\mathbf{a}(\theta_{m})-\sigma_{e}^{2}N_{t}|^{2}     \label{eqn:obj_2}   \\
	\text{s.t.}&  \quad  \sum_{k \in \mathcal{K}}C_{k} \leq R_{c,k}(\mathbf{P}, \delta, \sigma_{e}), \quad  k \in \mathcal{K} \label{eqn:common_rate_2} \\
	& \quad   \mathbf{c} \geq \mathbf{0} \label{eqn:common_rate_3} \\
	& \quad \mathrm{diag}\left( \mathbf{P}\mathbf{P}^{H}\right)+P(\delta)\mathbf{1}=\frac{P_{\mathrm{total}}\mathbf{1}}{N_{t}} \label{eqn:radar_power_1}  \\
	& \quad   \alpha > 0. \label{eqn:alpha} 
	\end{alignat}
	\label{eqn:problem2}
\end{subequations}
\hspace{-0.3cm} The rates $C_{k}$ and $R_{k}(\mathbf{P}, \delta, \sigma_{e})$ are the common and private rates of user-$k$, respectively, and $\mathbf{c}=[C_{1},C_{2}, \ldots\, C_{K}]^{T}$ is the vector of common rates. The rate $R_{c,k}(\mathbf{P}, \delta, \sigma_{e})$ denotes the total rate of the common stream at user-$k$. The parameter $\lambda$ in the objective function is the regularization parameter that performs weighting between the communication sum-rate and radar beampattern error.  $P_{\mathrm{total}}$ denotes the total transmit power budget to be shared between the transmit precoders and the DACs at each antenna. 
From the constraint \eqref{eqn:radar_power_1} and the expressions \eqref{eqn:delta} and \eqref{eqn:power}, the power allocated to the precoders is obtained as 
\begin{align}
    \mathrm{tr}(\mathbf{P}\mathbf{P}^{H})=P_{\mathrm{total}}-N_{t}P(\delta)=P_{\mathrm{total}}-2^{b}N_{t}P_{\mathrm{DAC}}.
\end{align}

Next, we discuss the ADMM approach to solve problem formulation \eqref{eqn:problem2}.
 
\begin{figure}[t!]
 \removelatexerror
  \begin{algorithm}[H]
	 \caption{ADMM-Based Algorithm}
		\label{alg:algorithm}
			$t \gets 0$, $\mathbf{v}^{0}_{r}$, $\mathbf{u}^{0}_{r}$, $\mathbf{d}^{0}$ \\
			\While{$||\mathbf{r}^{t}||_{2} \leq \epsilon$ $\mathrm{and}$ $||\mathbf{q}^{t}||_{2} \leq \epsilon$}{
			$\mathbf{v}^{t+1}_{r} \gets \argmin_{\mathbf{v}_{r}}\mathcal{L}_{\rho}(\mathbf{v}_{r},\mathbf{u}^{t}_{r},\mathbf{d}^{t})$ via WMMSE-AO algorithm\\
			$\mathbf{u}^{t+1}_{r} \gets \argmin_{\mathbf{u}_{r}}\mathcal{L}_{\rho}(\mathbf{v}^{t}_{r},\mathbf{u}_{r},\mathbf{d}^{t})$ via SDR algorithm\\
			$\mathbf{y}^{t+1} \gets \mathbf{y}^{t}+\rho\mathbf{D}_{pr}(\mathbf{v}^{t+1}_{r}-\mathbf{u}^{t+1}_{r})$\\
			$\mathbf{r}^{t+1} \gets \mathbf{D}_{pr}(\mathbf{v}^{t+1}_{r}-\mathbf{u}^{t+1}_{r})$\\
			$\mathbf{q}^{t+1} \gets \mathbf{D}_{pr}(\mathbf{u}^{t+1}_{r}-\mathbf{u}^{t}_{r})$\\
			$t \gets t + 1$\\
			}
  \end{algorithm}
  \vspace{-0.5cm}
\end{figure}
 
 \begin{figure}[t!]
	\hspace{0.2cm}\centerline{\includegraphics[width=3.9in,height=3.9in,keepaspectratio]{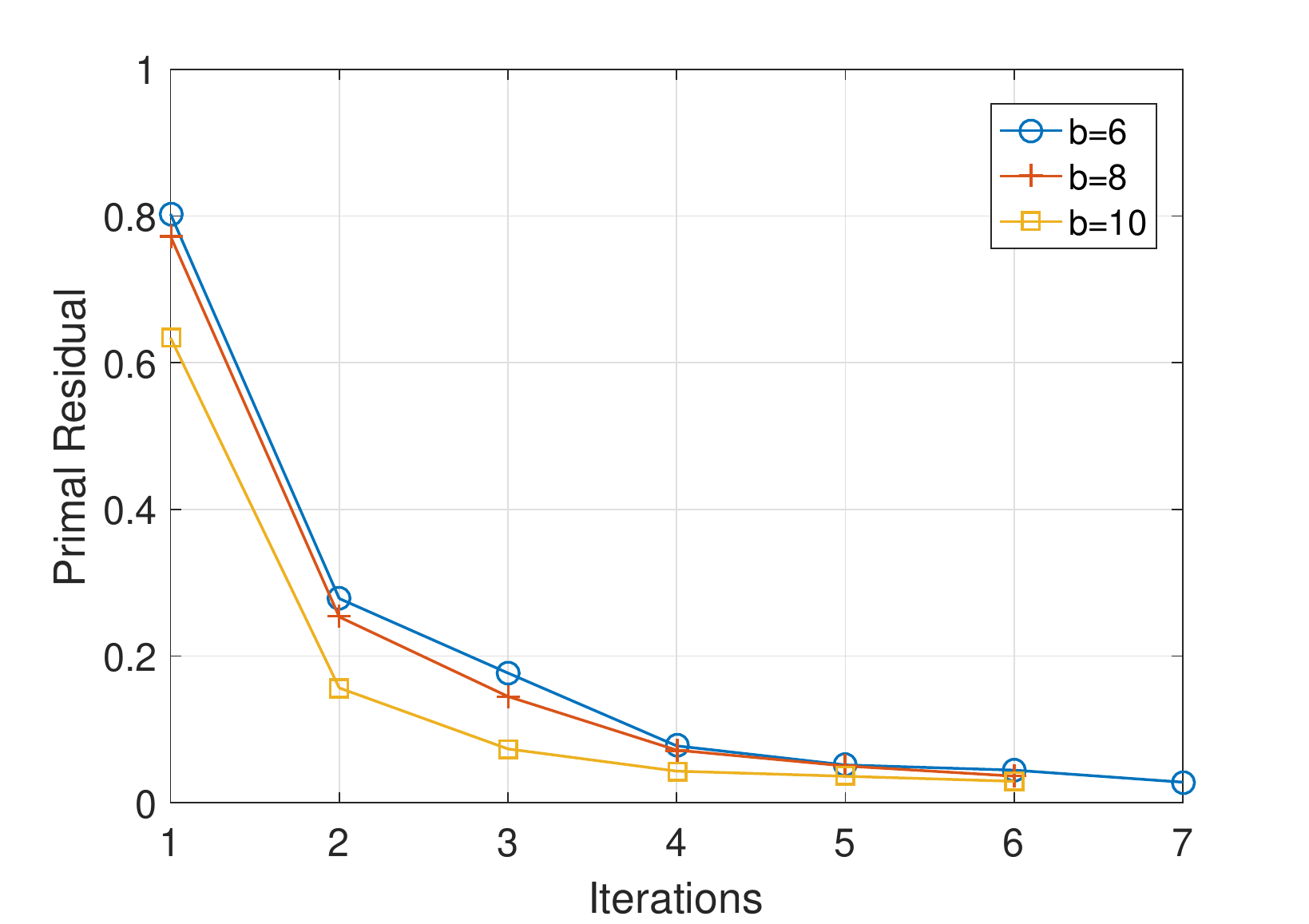}}
	\caption{Convergence behaviour of the algorithm}
	\label{fig:convergence}
\end{figure}

\begin{figure}[t!]
	\begin{subfigure}{.5\textwidth}
	\centerline{\includegraphics[width=3.5in,height=3.5in,keepaspectratio]{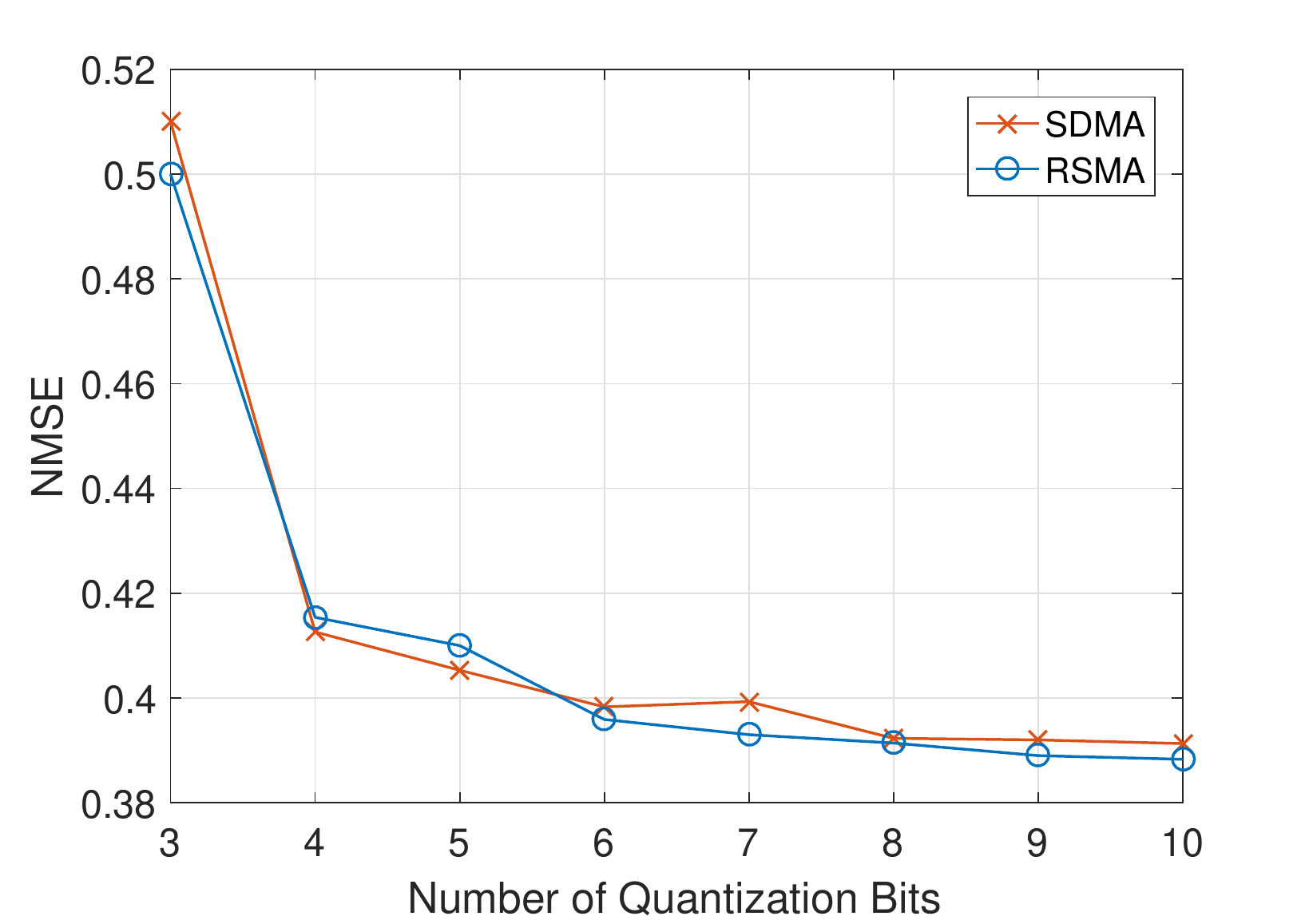}}
	\caption{NMSE vs. $b$.}
	\label{fig:nmsevsQ}
	\end{subfigure}
	\begin{subfigure}{.5\textwidth}
	\centerline{\includegraphics[width=3.7in,height=3.7in,keepaspectratio]{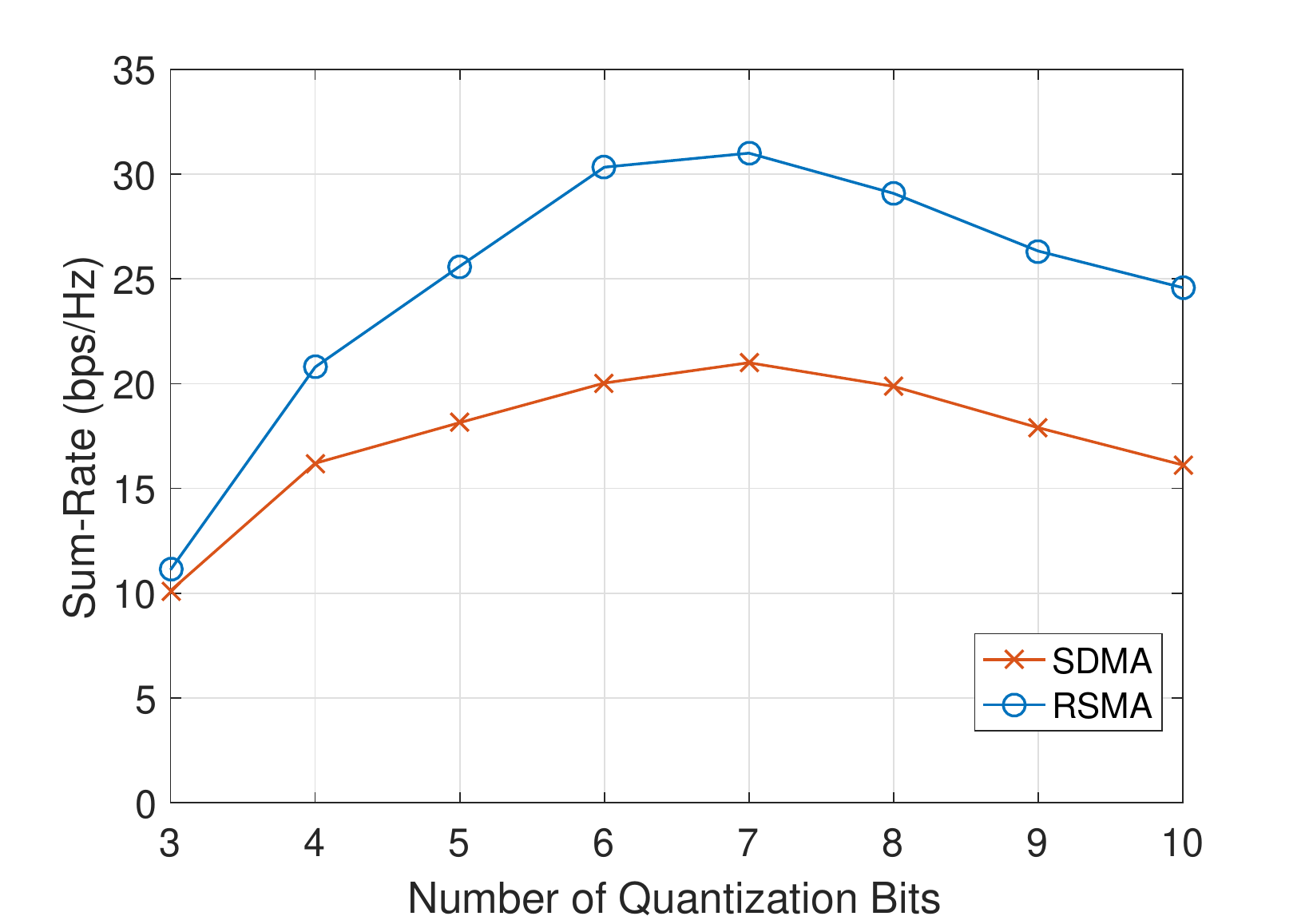}}
	\caption{Sum-rate vs. $b$.}
	\label{fig:srvsQ}
	\end{subfigure}
	\caption{Sum-rate and NMSE performance of RSMA and SDMA.}
\end{figure}

\subsection{ADMM-Based Solution} 
For a given number of bits $b$, or equivalently $\delta$, the optimization problem in \eqref{eqn:problem2} can be solved by the ADMM method \cite{boyd_2011}. Such a solution has been also discussed in the context of RSMA for JRC in \cite{xuICCW2020} without considering the effects of low resolution DACs in the system.
First, we reformulate the problem \eqref{eqn:problem2} into an equivalent form 
\begin{subequations}
	\begin{alignat}{3}
		\min_{\mathbf{v},\mathbf{u}}&     \quad  f_{c}(\mathbf{v})+g_{c}(\mathbf{v})+f_{r}(\mathbf{u})+g_{r}(\mathbf{u})     \label{eqn:obj_3}   \\
		\text{s.t.}&  \quad  \mathbf{D}_{p}(\mathbf{v}-\mathbf{u})=\mathbf{0},
	\end{alignat}
	\label{eqn:problem_admm}
\end{subequations}
\hspace{-0.3cm}where $\mathbf{v}=[\alpha, \mathbf{c}^{T}, \mathrm{vec}(\mathbf{P})^{T}]^{T}$, $\alpha \in \mathbb{R}_{++}$ and \mbox{$\mathbf{D}_{p}=[\mathbf{0}^{(K+1)N_{t} \times (K+1)}, \ \mathbf{I}_{(K+1)N_{t}}]$}. 
The vector $\mathbf{u}$ is introduced to split the problem according to the ADMM procedure. The function $f_{c}(\mathbf{v})$ is the objective function for sum-rate maximization expressed as a minimization problem and in terms of $\mathbf{v}$ as
\begin{align}
    f_{c}(\mathbf{v})=-\sum_{k \in \mathcal{K}}(\mathbf{e}_{k+1}^{T}\mathbf{v}+R_{k}(\mathbf{D}_{p}\mathbf{v}, \delta, \sigma_{e})),
\end{align}
where $\mathbf{e}_{k}$ is the $k$-th standard basis vector of length \mbox{$N_{t}(K+1)+K+1$}. 
Similarly, $f_{r}(\mathbf{u})$ is the objective function for radar beampattern error minimization expressed as
\begin{align}
    f_{r}(\mathbf{u})=\lambda\sum_{m=1}^{M}|\alpha &P_{d}(\theta_{m})-\delta^{2}\mathbf{a}^{H}(\theta_{m})(\mathbf{D}_{c}\mathbf{u}\mathbf{u}^{H}\mathbf{D}_{c}^{H}\nonumber \\
    &+\sum_{k \in \mathcal{K}}\mathbf{D}_{k}\mathbf{u}\mathbf{u}^{H}\mathbf{D}_{k}^{H})\mathbf{a}(\theta_{m})-\sigma_{e}^{2}N_{t}|^{2},\nonumber
\end{align}
where 
\begin{align}
    \mathbf{D}_{c}&=[\mathbf{0}^{N_{t}\times(K+1)}, \ \mathbf{I}_{N_{t}}, \ \mathbf{0}^{N_{t}\times(KN_{t})}], \nonumber \\
    \mathbf{D}_{k}&=[\mathbf{0}^{N_{t}\times(K+1+kN_{t})}, \ \mathbf{I}_{N_{t}}, \ \mathbf{0}^{N_{t}\times((K-k)N_{t})}]. \nonumber
\end{align}
The functions $g_{c}(\mathbf{v})$ and $g_{r}(\mathbf{u})$ are the feasible sets for the set of vectors $\mathbf{v}$ satisfying the constraints \eqref{eqn:common_rate_2} and \eqref{eqn:common_rate_3} and set of vectors $\mathbf{u}$ satisfying the constraints \eqref{eqn:radar_power_1} and \eqref{eqn:alpha}, respectively. 

We express the augmented Lagrangian function for the optimization problem \eqref{eqn:problem_admm} as
\begin{align}
    \mathcal{L}_{\rho}(\mathbf{v}_{r},\mathbf{u}_{r},\mathbf{d})&=f_{c}(\mathbf{v}_{r})+g_{c}(\mathbf{v}_{r})+f_{r}(\mathbf{u}_{r})+g_{r}(\mathbf{u}_{r})+ \nonumber \\
    &\mathbf{y}^{T}(\mathbf{D}_{pr}(\mathbf{v}_{r}-\mathbf{u}_{r}))+(\rho/2)||\mathbf{D}_{pr}(\mathbf{v}_{r}-\mathbf{u}_{r})||^{2}_{2}.\nonumber
\end{align}
\noindent The vectors $\mathbf{u}_{r}$ and $\mathbf{v}_{r}$ are real valued vectors consisting of the real and imaginary parts of the corresponding complex-valued vectors to match with the definition of ADMM \cite{boyd_2011}, and are defined as \mbox{$\mathbf{v}_{r}=[\mathfrak{R}(\mathbf{v}^{T}), \mathfrak{I}(\mathbf{v}^{T})]^{T}$}, \mbox{$\mathbf{u}_{r}=[\mathfrak{R}(\mathbf{u}^{T}), \mathfrak{I}(\mathbf{u}^{T})]^{T}$} and \mbox{$\mathbf{D}_{pr}=\mathrm{Diag}(\mathbf{D}_{p}, \mathbf{D}_{p})$}.
The updates of the iterative ADMM procedure steps are written as
\begin{align}
    \mathbf{v}^{t+1}_{r}&=\argmin_{\mathbf{v}_{r}}\mathcal{L}_{\rho}(\mathbf{v}_{r},\mathbf{u}^{t}_{r},\mathbf{d}^{t})  \\
    \mathbf{u}^{t+1}_{r}&=\argmin_{\mathbf{u}_{r}}\mathcal{L}_{\rho}(\mathbf{v}^{t}_{r},\mathbf{u}_{r},\mathbf{d}^{t}) \\
    \mathbf{y}^{t+1}&=\mathbf{y}^{t}+\rho\mathbf{D}_{pr}(\mathbf{v}^{t+1}_{r}-\mathbf{u}^{t+1}_{r}).
\end{align}

\noindent The ADMM-based algorithm to solve the problem formulation \eqref{eqn:problem_admm} is given in Algorithm~\ref{alg:algorithm}. The algorithm benefits from the Weighted Minimum Mean Square Error - Alternating Optimization (WMMSE-AO) based algorithm in \cite{clerckxTC2016} to solve the sum-rate maximization problem and Semi-Definite Relaxation (SDR) method to solve the radar beampattern error minimization problem \cite{ma_2010}. In the next section, we evaluate the performance the described algorithm by simulation results.

\begin{figure*}[t!]
	\begin{subfigure}{.5\textwidth}
	\centerline{\includegraphics[width=3.3in,height=3.3in,keepaspectratio]{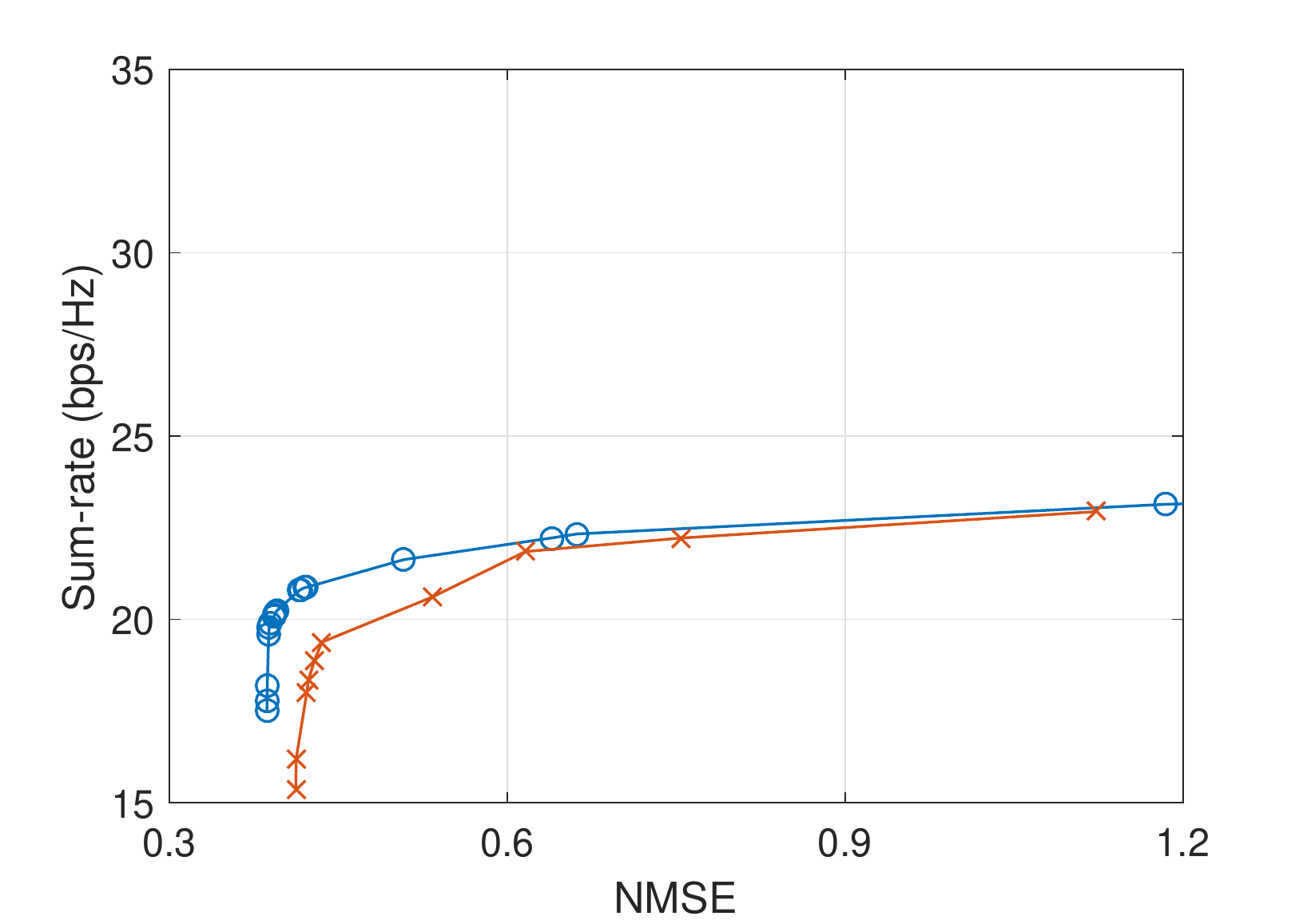}}
	\caption{$b=4$.}
	\label{fig:Qbits4_SRvsNMSE}
	\end{subfigure}
	\begin{subfigure}{.5\textwidth}
	\centerline{\includegraphics[width=3.3in,height=3.3in,keepaspectratio]{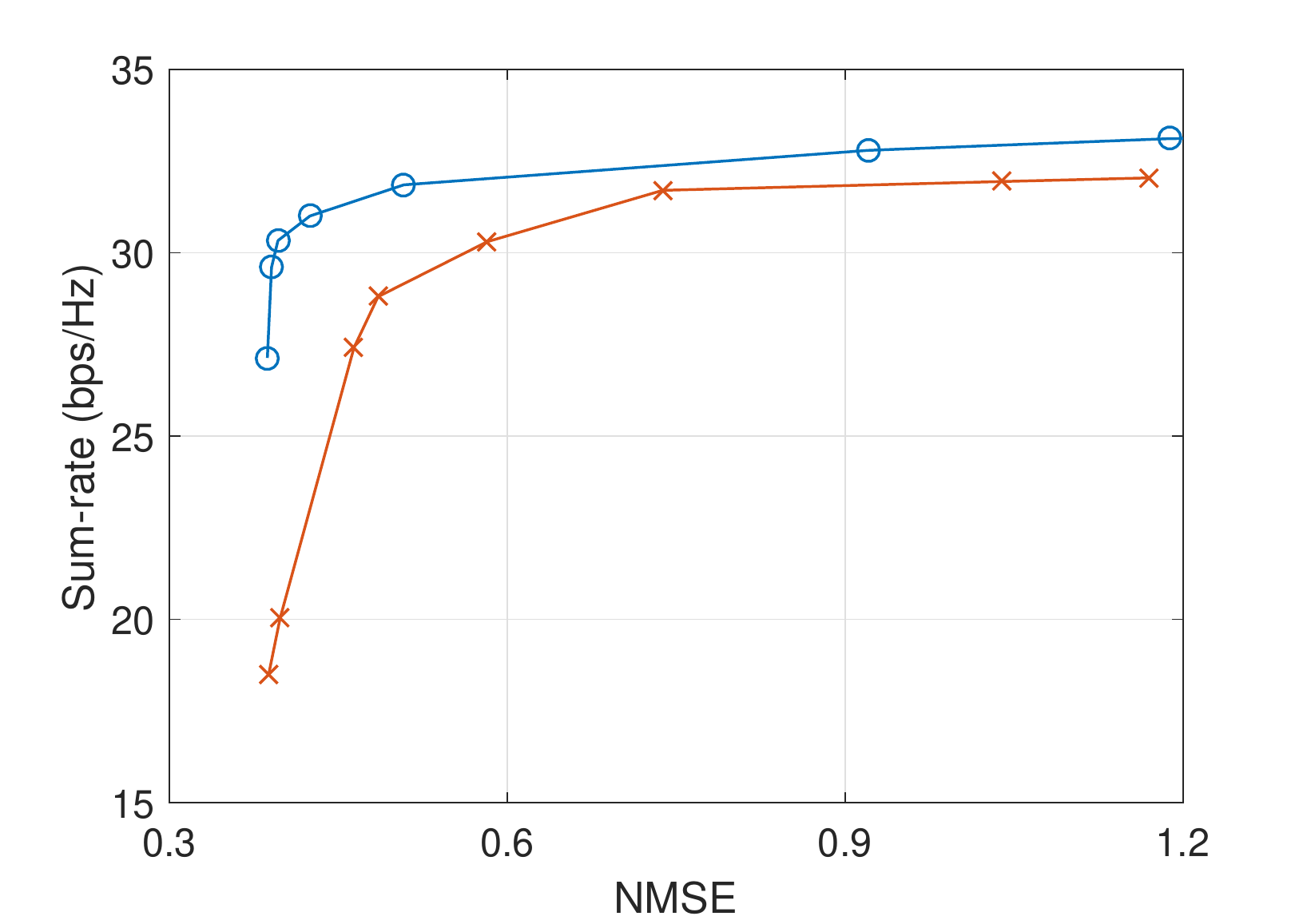}}
	\caption{$b=6$.}
	\label{fig:Qbits6_SRvsNMSE}
	\end{subfigure}
	\newline
	\begin{subfigure}{.5\textwidth}
	\centerline{\includegraphics[width=3.3in,height=3.3in,keepaspectratio]{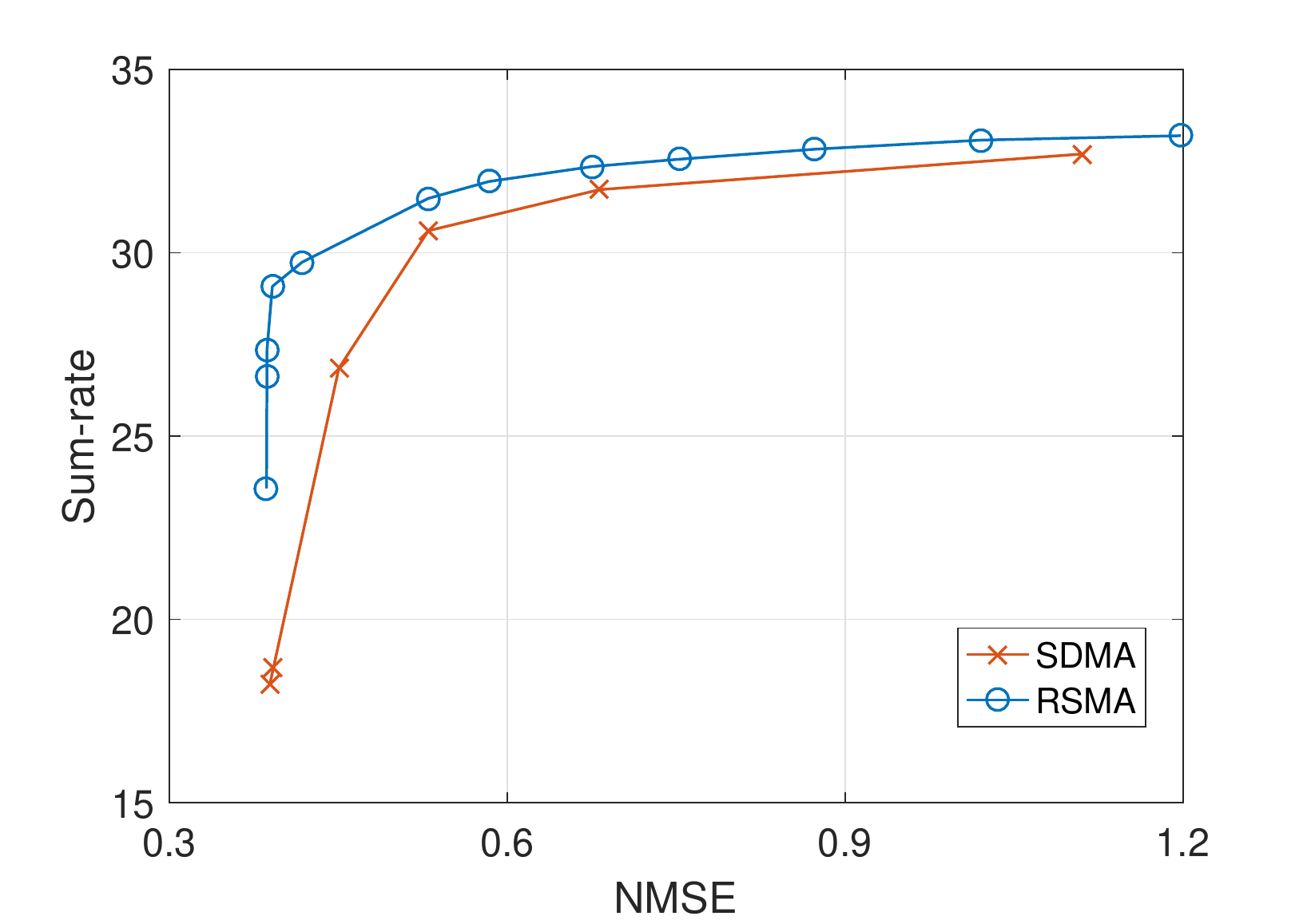}}
	\caption{$b=8$.}
	\label{fig:Qbits8_SRvsNMSE}
	\end{subfigure}
	\begin{subfigure}{.5\textwidth}
	\centerline{\includegraphics[width=3.3in,height=3.3in,keepaspectratio]{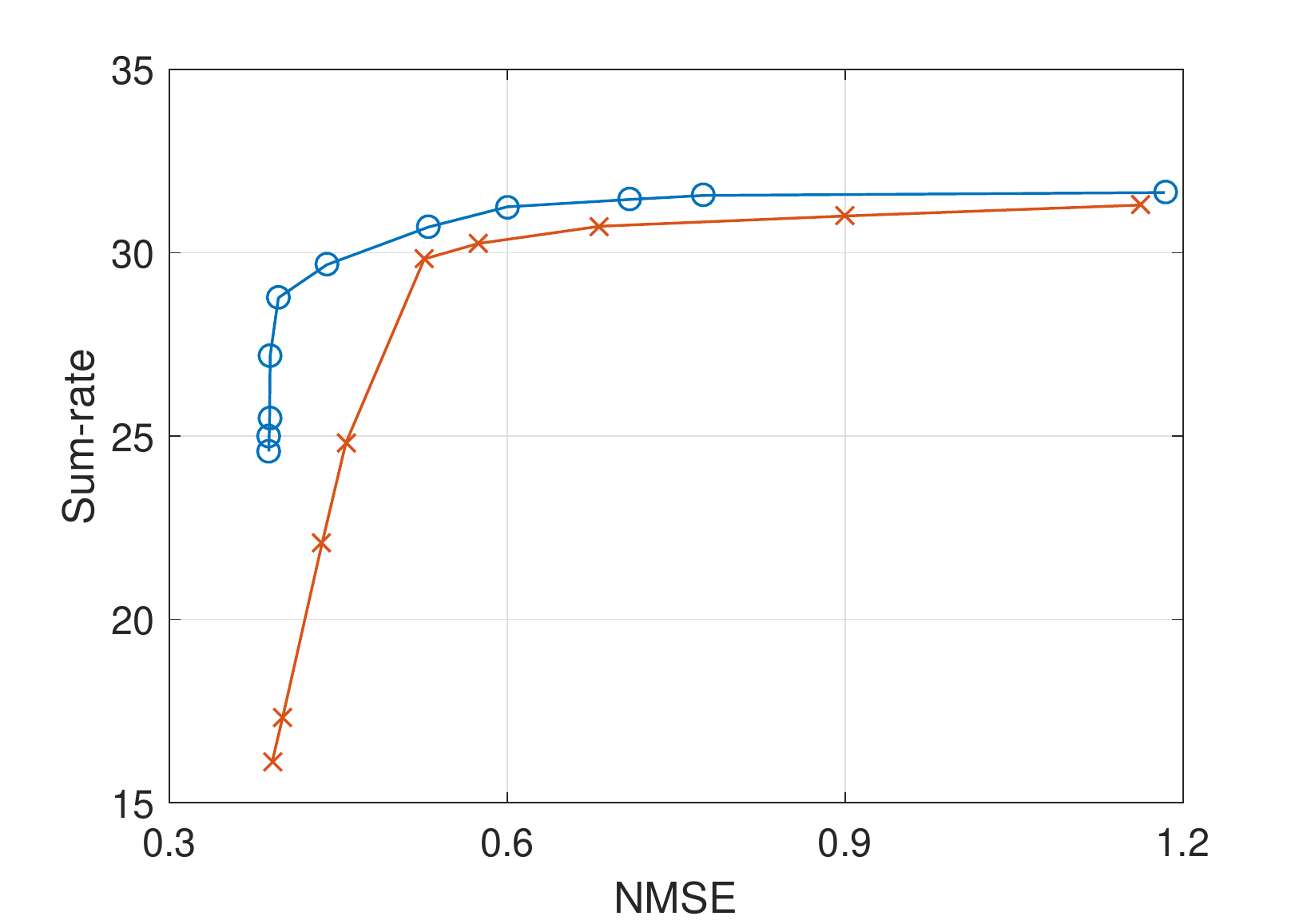}}
	\caption{$b=10$.}
	\label{fig:Qbits10_SRvsNMSE}
	\end{subfigure}
	\caption{NMSE vs. sum-rate for varying $b$.}
	\label{fig:SRvsNMSE}
\vspace{-0.5cm}
\end{figure*}

\section{Simulation Results}

In this section, we present simulation results to evaluate the performance of RSMA in a JRC system under DAC quantization, and compare it with SDMA. We note that the optimal precoders for SDMA are obtained by turning off the common stream in the optimization problem. We set the system and antenna settings as $K=2$, $N_{t}=4$, $d=0.5$. 
The total transmit budget is $P_{total}=1$W, the DAC power consumption coefficient is $P_\textrm{DAC}=100\mu$W and the noise power at the receiver is $N_{0}=10\mu$W. The radar target is located at the $0$-degree azimuth direction. We generate single realizations of Rayleigh fading channels independently for each user and perform the simulations over these specific realizations. 

The convergence behaviour of the Algorithm~\ref{alg:algorithm} is demonstrated in Figure~\ref{fig:convergence} in terms of the primary residual $||\mathbf{r}^{t}||_{2}$ for $\lambda=1$ and varying $b$. As observed from the figure, the algorithm converges for the investigated $b$ values, with a different number of iterations for each $b$.

We investigate the Normalized Mean Square Error (NMSE) of the radar beampattern to evaluate the radar performance, which is defined as
\begin{align}
     \mathrm{NMSE}\hspace{-0.1cm}=\hspace{-0.1cm}\frac{\sum_{m=1}^{M}\hspace{-0.1cm}|\alpha P_{d}(\theta_{m})-\delta^{2}\mathbf{a}^{H}(\theta_{m}) \mathbf{P}\mathbf{P}^{H}\mathbf{a}(\theta_{m})-\sigma_{e}^{2}N_{t}|^{2}}{\sum_{m=1}^{M}|\alpha P_{d}(\theta_{m})|^{2}}. \nonumber
\end{align}

Figures~\ref{fig:nmsevsQ}~and~\ref{fig:srvsQ} present the sum-rate and NMSE performance of RSMA and SDMA for $\lambda=10$. 
As observed from Figure~\ref{fig:nmsevsQ}, the NMSE of the radar waveform retains a monotonic decreasing behaviour with increasing $b$.
On the other hand, the sum rate performance in Figure~\ref{fig:srvsQ} do not follow a monotonic behaviour. Such phenomenon is explained as follows. 
The sum-rate depends on the SINR of the streams, as given in \eqref{eqn:sinr_1} and \eqref{eqn:sinr_2}, and the power allocated to the precoders as in the constraint \eqref{eqn:radar_power_1}. As discussed in Section~\ref{sec:problem}, the power allocated to the precoders decreases with increasing $b$. 
Hence at small values of $b$, the system has large quantization error but large power for the precoders, while at large values of $b$, the quantization error decreases along with the power of the precoders.
Consequently, the highest SINR is achieved at an intermediate value of $b$, at which both RSMA and SDMA achieve highest sum-rate performance. 

As observed from Figures~\ref{fig:nmsevsQ}~and~\ref{fig:srvsQ}, RSMA achieves significantly higher sum-rate than SDMA for similar NMSE values. The improved performance of RSMA is a result of its ability to manage the interference resulting from the radar beampattern under the strict per antenna power constraint \eqref{eqn:radar_power_1}. The SINR expressions \eqref{eqn:sinr_1} and \eqref{eqn:sinr_2} show that the quantization error affects the sum-rate performance only by altering the operating SINR of the system and does not yield an additional source of manageable interference, since the additive random quantization noise $\boldsymbol{\epsilon}$ is independent of the transmitted signal. 
The performance gain of RSMA varies with $b$ as the effective SINR is dependent on the number of quantization bits and the gain achieved by RSMA over SDMA increases with SINR \cite{clerckx2018, clerckxWCL2020}. 
For small values of $b$, the quantization error is large, resulting in low effective SINR and smaller gain by RSMA. As $b$ becomes sufficiently large, the gain of RSMA over SDMA increases. When $b$ increases beyond a certain value, the effective SINR starts to drop again due to the increasing DAC power consumption and decreasing transmit power, resulting in a slight decrease in the gain.

Finally, we investigate the sum-rate performance with respect to NMSE for varying number of quantization bits in Figures~\ref{fig:Qbits4_SRvsNMSE}-\ref{fig:Qbits10_SRvsNMSE}.
As observed from the figures, RSMA achieves significantly higher sum-rate compared to SDMA in the considered NMSE regions. The performance gain varies with the number of quantization bits, as also observed in Figure~\ref{fig:srvsQ}.

\section{Conclusion}
In this work, we studied RSMA and SDMA for JRC systems under DAC quantization errors. We designed optimal precoders which maximize the sum-rate and minimize the radar beampattern error jointly for given number of quantization bits and total transmit power budget. We analyzed the trade-off induced by increasing the number of quantization bits to improve the quality of the precoders, which in turn increases the power consumption of DACs in the system and decreases the transmit power due to the considered total transmit power budget. We demonstrate by simulations that the maximum sum-rate is achieved by a number of quantization bits that is less than the maximum number allowed by the transmit power budget. We further show that RSMA achieves a significant performance gain in terms of sum-rate with respect to SDMA for similar achieved radar beampattern NMSE. We conclude that RSMA outperforms SDMA in JRC systems with low-resolution DACs and a total transmit budget.

\section*{Acknowledgment}
This work was supported by the Engineering and Physical Sciences Research Council of the UK (EPSRC) Grant numbers EP/S026622/1 and EP/S026657/1, and the UK MOD University Defence Research Collaboration (UDRC) in Signal Processing.

\end{document}